\documentclass[aps,prl,reprint,groupedaddress]{revtex4-1}
\usepackage{graphicx} 
\usepackage{dcolumn} 
\usepackage{bm}
\usepackage{hyperref}

\begin{document}
\title{ Shedding light on the pion production in heavy-ion collisions and application into the neutron star matter properties }

\author{Heng-Jin Liu$^{1}$}
\author{Ban Zhang$^{1}$}
\author{Yuan-Qing Guo$^{1}$}
\author{Hui-Gan Cheng$^{1}$}
\author{Si-Na Wei$^{2}$}
\author{Zhao-Qing Feng$^{1}$}
\email{Corresponding author: fengzhq@scut.edu.cn}

\affiliation{$^{1}$School of Physics and Optoelectronics, South China University of Technology, Guangzhou 510640, China}
\affiliation{$^{2}$School of Mathematics and Physics, Guangxi Minzu University, Nanning 530006, China}

\date{\today}

\begin{abstract}

Within the framework of the quantum molecular dynamics transport model, the pion production and constraint of the high-density symmetry energy in heavy-ion collisions near threshold energy have been thoroughly investigated. The energy conservation in the decay of resonances and reabsorption of pions as well as in the inelastic nucleon-nucleon and nucleon-resonance collisions are taken into account. The isospin diffusion in the low-density region (0.2$\rho_{0}$ - 0.8$\rho_{0}$) and high-density region (1.2$\rho_{0}$ - 1.8$\rho_{0}$) is investigated by analyzing the spectra of neutron/proton and $\pi^{-}/\pi^{+}$ ratios in the isotopic reactions of $^{132}$Sn + $^{124}$Sn and $^{108}$Sn + $^{112}$Sn at the incident energy of 270 MeV/nucleon, in which the symmetry energy manifests the opposite effect in the different density domain. The controversial conclusion of the $\pi^{-}/\pi^{+}$ ratio for constraining the high-density symmetry energy by different transport models with the FOPI data has been clarified. A soft symmetry energy with the slope parameter of $L(\rho_{0}) = 42\pm 25$ MeV by using the standard error analysis within the range of $1\sigma$ is obtained by analyzing the experimental data from the S$\pi$RIT collaboration. The neutron stars with the maximal mass of 2 $M_{\odot}$ and radius of 11-13 km are obtained with the constrained symmetry energy.

\begin{description}
\item[PACS number(s)]
21.65.Ef, 21.65.Jk, 24.10.Lx   
\end{description}
\end{abstract}

\maketitle

\section{I. Introduction}

Heavy-ion collisions at intermediate energies under the extreme conditions, such as high density, high temperature, and high isospin asymmetry, provide a good environment for exploring the properties of dense nuclear matter. The particle production in heavy-ion collisions brings the information of the hot and compressed nuclear matter. Pion meson has been attracted attention for probing the dense matter, which was predicted by Yukawa as a propagator of the strong interaction \cite{Yu35} and later discovered by observing the cosmic rays in experiments \cite{La47}. Nowadays, pions might be created in laboratories via different reactions, such as heavy-ion collisions, photon, lepton and hadron induced reactions, electron-positron colliding etc. Particles produced in the GeV energy range manifest the information of the high-density nuclear matter and might be modified by the nuclear medium \cite{Ki97}. The nuclear equation of state (EOS) is expressed with the energy per nucleon as E($\rho,\delta$) = E($\rho,\delta=0$) + $E_{sym}(\rho)\delta^2$ + $\mathcal{O}(\delta^4)$ in terms of baryon density $\rho$ = $\rho_n$+$\rho_p$ and relative neutron excess $\delta$ = ($\rho_n - \rho_p$)/($\rho_n + \rho_p$), where $E_{sym}(\rho)$ is the symmetry energy \cite{Bo91,Ba05,Li08}. The symmetry energy at the subsaturation density has important application in understanding the structure of weakly bound nuclei, nucleon-nucleon correlation, pasta structure of neutron star etc, which was extensively investigated by different approaches, such as the Pygmy dipole resonance, heavy-ion collision, fast fission, electron-nucleus scattering etc \cite{Li97,Ch05,Ts08,Ca15,Su10,Fi14,Fe16,Fe18,Zh17a,Ad21}. The high-density symmetry energy is related to the issues of compact stars, such as the phase transition, binary neutron star merging, tidal deformation etc \cite{St05,Fr07,Di11,Ba13,Oe17,Hu22}, which is still not well understood up to now. To extracting the density dependence of symmetry energy, new experiments are being carried out at the facility in the world, such as Radioactive Isotope Beam Facility (RIBF) in Japan \cite{Sa18}, Rare Isotope Science Project in Korea (RAON) \cite{Ts13}, Facility for Rare Isotope Beams (FRIB) in the USA \cite{Os19}, the Cooling Storage Ring (CSR) and the High Intensity Accelerator Facility (HIAF) in China \cite{Ya13}.

There has been performed a number of experimental data for the pion production in heavy-ion collisions, such as the BEVALAC with the Berkeley Streamer Chamber \cite{Ha85,Ha87}, the DIOGENE at the Saturne synchrotron in Saclay \cite{Al87}, the FOPI collaboration at the beam energies from 0.4\emph{A} to 1.5\emph{A} GeV at Gesellschaft f\"{u}r Schwerionenforschung \cite{Re07}, HIRFL-CSR in Lanzhou \cite{Wa19}, the S${\pi}$RIT collaboration at RIKEN for the reactions of $^{132}$Sn + $^{124}$Sn and $^{108}$Sn + $^{112}$Sn at 270\emph{A} MeV \cite{Es21}. The opposite conclusion for constraining the high-density symmetry energy was obtained by analyzing the FOPI data of the $\pi^{-}/\pi^{+}$ excitation function with the different transport models \cite{Xi09,Fe10a}, namely the 'pion puzzle'. At the near threshold energy, the pion mesons are mainly produced via the decay of $\Delta$(1232) resonance. The pion dynamics is modified in the nuclear environment, i.e., the pion-nucleon potential, elementary production cross section, threshold energy, decay width of resonance, reabsorption process via the reaction $\pi$N$\leftrightarrow \Delta$ etc. The influence of the $\pi$ potential on the pion dynamics in heavy-ion collisions has been investigated via transport models \cite{Xi93,Fu97,Fe10,Ho14,Fe15,Gu15,So15,Co17}. Recently, the observable uncertainties in transport models for heavy-ion collisions are investigated by the transport model evaluation project (TMEP) \cite{Ak19,He22}. More sophisticated investigation of the pion-nucleon potential is still necessary, in particular, distinguishing the isospin effect, manifesting the momentum and density dependence and reproducing the pion-nucleus scattering data. Both the pion-nucleon potential and stiffness of symmetry energy influence the $\pi^-$/$\pi^+$ momentum spectra in heavy-ion collisions.

In this work, the pion production in the isotopic reactions near threshold energy and high-density symmetry energy are to be investigated with the Lanzhou quantum molecular dynamics (LQMD) transport model. The article is organized as follows. In Sec. 2, we briefly introduce the theoretical method and some improvements. The calculated results and comparison with the available data from the S${\pi}$RIT collaboration are shown in Sec. 3. A summary perspective on the future experiments is given in Sec. 4.

\section{II. Theoretical approach }

\begin{figure}
    \includegraphics[height=8cm, width=8cm]{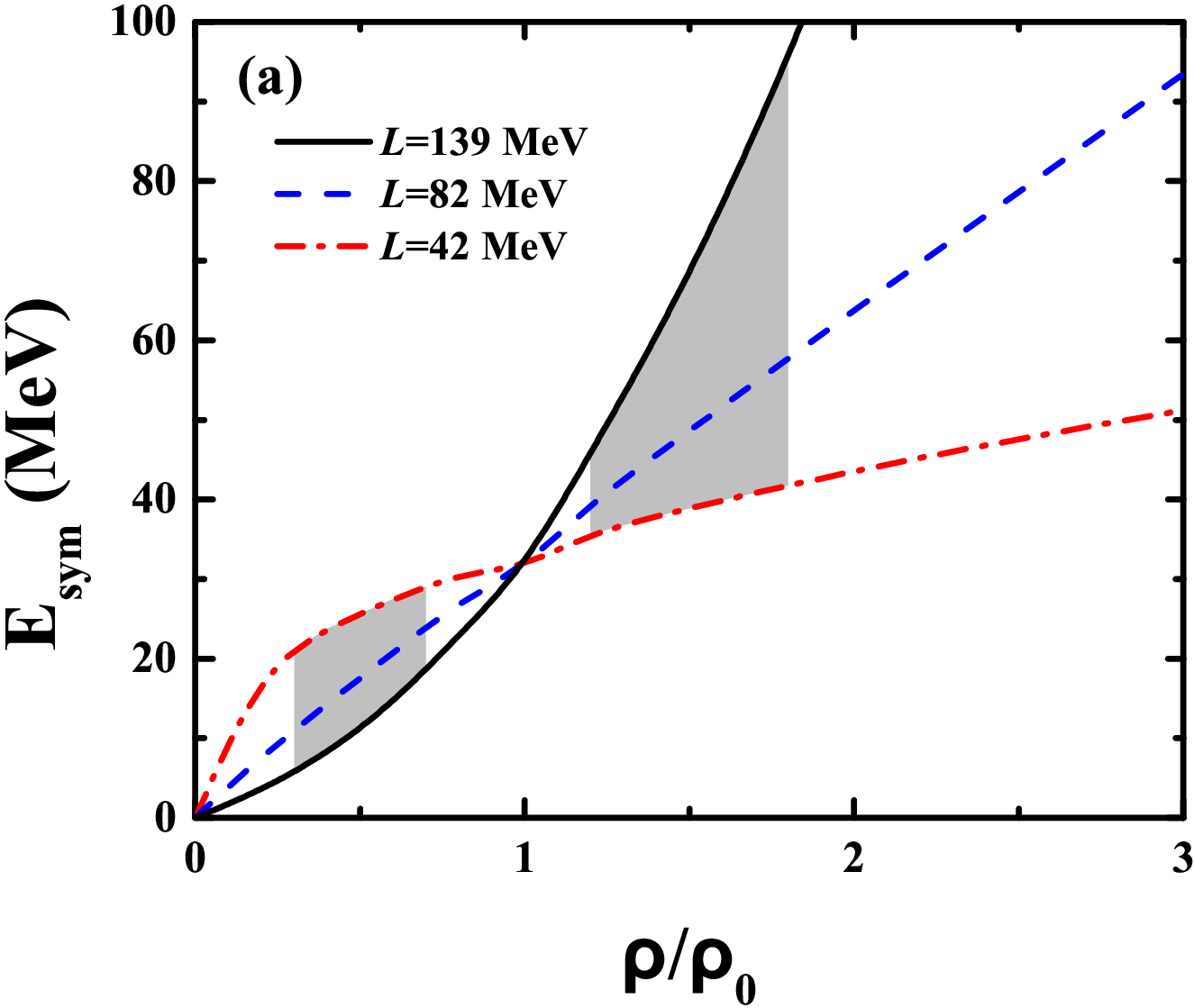}
    \includegraphics[height=8cm, width=8cm]{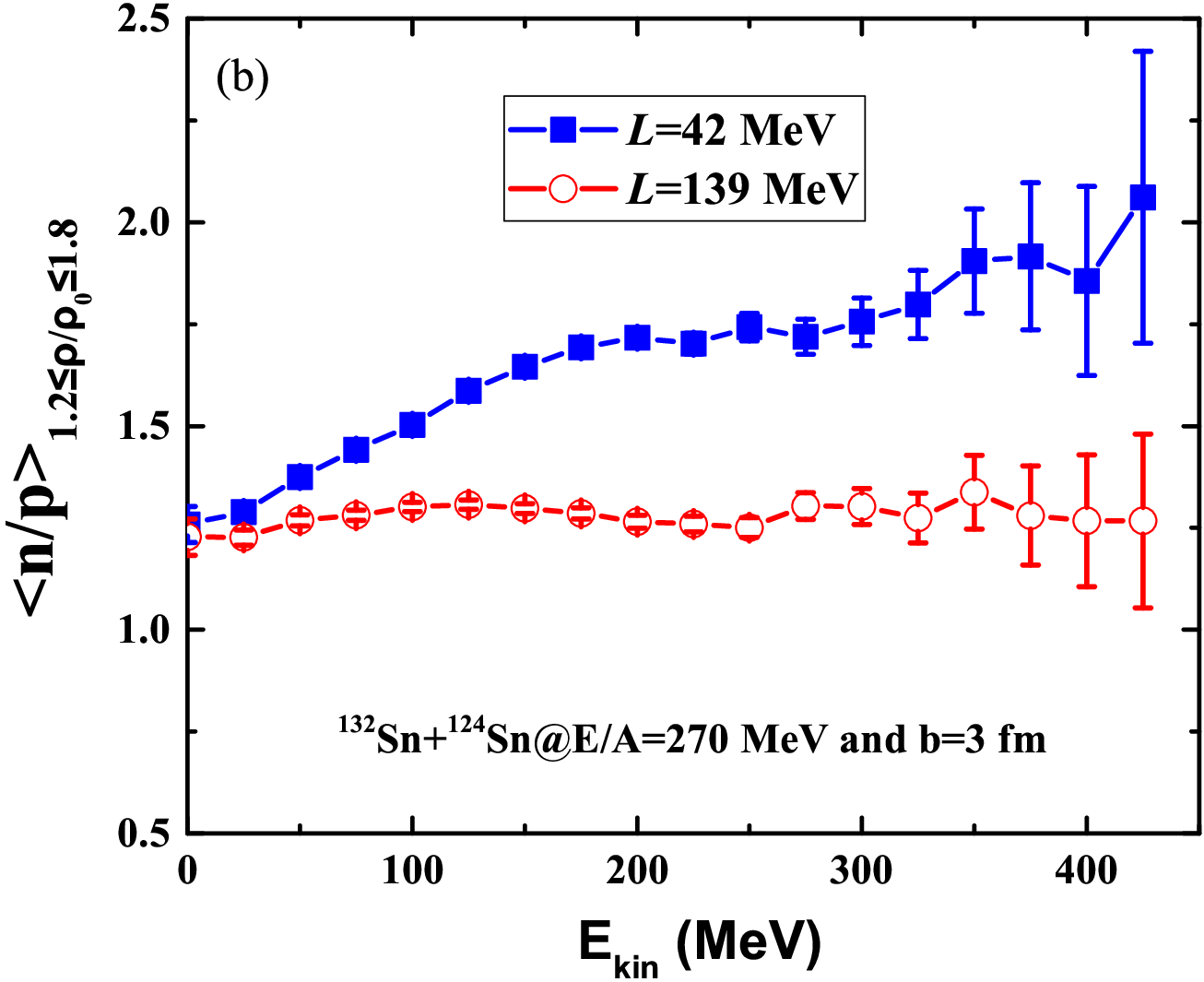}
    \caption{(a) The density dependence of nuclear symmetry energy with different stiffness and (b) the neutron/proton ratio at the supra-saturation densities 1.2$\leq\rho/\rho_{0}\leq$1.8 with the temporal evolutions in collisions of $^{132}$Sn+$^{124}$Sn at the beam energy of 270 MeV/nucleon and with the impact parameter of b=3 fm.}
\label{fig:1}
\end{figure}

In the LQMD transport model, the production of resonances, hyperons and mesons is coupled in the reactions of meson-baryon and baryon-baryon collisions, which has been used for the nuclear dynamics in heavy-ion collisions and hadron induced reactions \cite{Fe18,Fe11,Fe23,We24}. The temporal evolutions of nucleons and nucleonic resonances are described by Hamilton's equations of motion under the self-consistently generated two-body and three-body potentials with the Skyrme effective interaction. The symmetry energy is composed of three parts, namely, the kinetic energy from Fermi motion, the local density-dependent interaction, and the momentum-dependent potential, which reads as
\begin{equation}
E_{sym}(\rho)=\frac{1}{3}\frac{\hbar^{2} }{2m} (\frac{3}{2}\pi ^{2}\rho  )^{2/3} +E^{loc}_{sym}(\rho)+E^{mom}_{sym}(\rho).
\end{equation}
The stiffness of symmetry energy is adjusted by
\begin{equation}
E_{sym}^{loc}\left ( \rho  \right ) = \frac{1}{2} C_{sym}\left ( \rho /\rho _{0}  \right )^{\gamma _{s} }.
\end{equation}
The parameter $C_{sym}$ is 52.5 MeV, and the stiffness parameter $\gamma_s$ is adjusted for getting the density dependence of symmetry energy, e.g., the values of 0.3, 1, and 2 being the soft, linear, and hard symmetry energy, corresponding to the slope parameters $[L(\rho _{0}) = 3\rho_{0} dE_{sym}(\rho)/d\rho|_{\rho=\rho_{0}}]$ of 42, 82, and 139 MeV, respectively. As shown in Fig. \ref{fig:1}(a), it can been seen that in all cases of symmetry energy at the saturation density ($\rho_0$ = 0.16 fm$^{-3}$) with the value of 31.5 MeV. The different slope of symmetry energy influences the neutron/proton ratio and also the mass-radius relation of neutron star. The hard symmetry energy (L=139 MeV) results in the more repulsive force for neutrons in the neutron-rich matter and consequently leads to the lower neutron/proton ratio in the high-density domain and to the larger free neutron/proton ratio. The symmetry energy dominates the isospin diffusion in heavy-ion collisions and leads to the isospin density difference. The larger neutron/proton ratio in the nuclear matter with the density regime of 1.2$\leq\rho/\rho_{0}\leq$1.8 with the soft symmetry energy in the temporal evolution of $^{132}$Sn+$^{124}$Sn is obtained as shown in Fig. \ref{fig:1}(b), in particular in the regime of kinetic energy above 100 MeV. The supra-saturation densities 1.2$\leq\rho/\rho_{0}\leq$1.8 of protons and neutrons are accumulated from the averaged temporal evolution and judged from the local density formed in heavy-ion collisions. The stiffness of symmetry energy might be constrained by the isospin observables produced different density region in heavy-ion collisions.

The reaction channels for the pion production contributed from the resonance decay and the direct process, such as $\Delta$(1232), $N^{\ast}$(1440), $N^{\ast}$(1535) etc, are included in the model as follows \cite{Fe18}
\begin{eqnarray}
&& NN \leftrightarrow N\triangle, \quad  NN \leftrightarrow NN^{\ast}, \quad  NN
\leftrightarrow \triangle\triangle,  \nonumber \\
&& \Delta \leftrightarrow N\pi,  N^{\ast} \leftrightarrow N\pi,  NN \leftrightarrow NN\pi  (\texttt{s-state}),
\end{eqnarray}
Here the nucleon-nucleon (resonance) and pion-nucleon collisions are treated as the stochastic and isotropic scattering in the temporal evolution of reaction system. The momentum-dependent decay widths are implemented into the model for the resonances of $\Delta$(1232) and $N^{\ast}$(1440) and the elementary cross sections are taken from the parameterized formula calculated by the one-boson exchange model \cite{Hu94}. The constant width of $\Gamma$=150 MeV for the $N^{\ast}$(1535) decay is used in the calculation. The elastic scatterings in nucleon-nucleon, nucleon-resonance ($NR\rightarrow NR$) and resonance-resonance ($RR\rightarrow RR$) collisions and inelastic collisions of nucleon-resonance ($NR\rightarrow NN$, $NR\rightarrow NR^{\prime}$) and resonance-resonance channels ($RR\rightarrow NN$, $RR\rightarrow NR$, $RR\rightarrow RR^{\prime}$, $R$ and $R^{\prime}$ being different resonances), have been included in the model. The direct process $ NN \leftrightarrow NN\pi  (\texttt{s-state})$ roughly contributes 15$\%$ pion yields and the cross sections are taken as the same with the Giessen Boltzmann-Uehling-Uhlenbeck (GiBUU) transport model \cite{Bu12}.

The transportation of pion in nuclear medium is also described by the Hamiltonian equation of motion as
\begin{equation}
 H_{M} = \sum_{i=1}^{N_{M} }  \left[ V_{i}^{Coul} +  \omega \left (\textbf{p}_{i},\rho _{i}   \right)  \right].
\end{equation}
The Coulomb interaction is given by
\begin{equation}
    V_{i}^{Coul}= \sum_{j=1}^{N_{B} } \frac{e_{i}e_{j} }{r_{ij} }
\end{equation}
with $r_{ij}=|\textbf{r}_{i} - \textbf{r}_{j}|$. Here the $N_M$ and $N_B$ are the total numbers of mesons and baryons including charged resonances, respectively. It should be noticed that the pion meson is taken as the point particle and the Coulomb interaction between mesons is neglected owing to the limited numbers in comparison with the baryons.

The pion energy in the nuclear medium is composed of the isoscalar and isovector contributions as
\begin{equation}
    \omega _{\pi }  \left( \textbf{p}_{i} ,\rho _{i}  \right) = \omega _{isoscalar} \left(\textbf{p}_{i} ,\rho _{i}  \right) + C_{\pi }\tau _{z}\delta \left (\rho /\rho _{0}  \right )^{\gamma _{\pi } }.
\end{equation}
Here the isovector coefficient $C_{\pi}$ = $\rho_{0}\hbar^{3}/(4f^{2}_{\pi})$ = 36 MeV, isospin asymmetry $\delta$=($\rho_n - \rho_p$)/($\rho_n + \rho_p$ ) and isospin splitting parameter $\gamma_{\pi}$=2. The isospin quantities are set to be $\tau _{z}$= -1, 0, and 1 for $\pi^{+}$, $\pi^{0}$, and $\pi^{-}$, respectively \cite{Fe15}. The isoscalar part $\omega _{isoscalar} $ is estimated by the $\Delta$-hole model \cite{Br75,Fr81}.

 The energy balance in the decay of resonances and reabsorption of pions in nuclear medium is satisfied by the relation $R \leftrightarrow N\pi $ ($R$ being the resonance) as
 \begin{eqnarray}
&& \sqrt{m_{R}^{2}+\textbf{p}_{R}^{2}} + U_{R}(\rho,\delta,\textbf{p}_{R})  = \sqrt{m_{N}^{2} + \left (\textbf{p}_{R} - \textbf{p}_{\pi } \right )^{2} }         \nonumber \\
&& + U_{N}(\rho,\delta,\textbf{p})  + \omega _{\pi} \left (\textbf{p}_{\pi },\rho  \right)+V_{\pi N}^{Coul}.
 \end{eqnarray}
The $\textbf{p}_{R}$ and $\textbf{p}_{\pi}$ are the momenta of resonance and pion, respectively. The term $V_{\pi N}^{Coul}$ has the contribution only for the charged pair channels of $\triangle^{0}\leftrightarrow  \pi^{-}+p$ and $\triangle^{++}\leftrightarrow  \pi^{+}+p$, and no effect for the channels associated with the $\pi^{0}$ and neutron production. The optical potential can be evaluated from the in-medium energy $V_{\pi}^{opt}(\textbf{p},\rho) = \omega_{\pi} (\textbf{p},\rho) - (m_{\pi}^{2}+\textbf{p}^{2})^{1/2}$. The $U_{R}$ and $U_{N}$ are the singe-particle potentials for resonance and nucleon, respectively. The vacuum spectral function for the resonance production and decay is used in this work. Similarly, the energy conservation is also treated for the resonance production and the direct s-state creation, e.g., the collision with the local baryon density $\rho$ and the isospin asymmetry $\delta$ as
 \begin{eqnarray}
&& \sqrt{m_{N}^{2} + \textbf{p}_{1}^{2}} + U_{N}(\rho,\delta,\textbf{p}_{1}) + \sqrt{m_{N}^{2}+\textbf{p}_{2}^{2}} + U_{N}(\rho,\delta,\textbf{p}_{2})  =  \nonumber \\
&& \sqrt{m_{R}^{2}+\textbf{p}_{R}^{2}} + U_{R}(\rho,\delta,\textbf{p}_{R}) + \sqrt{m_{N}^{2}+{\textbf{p}}^{\prime 2}} + U_{N}(\rho,\delta,\textbf{p}^{\prime})
 \end{eqnarray}
for the channel $NN \leftrightarrow NR$ with the momentum conservation relation of $ \textbf{p}_{1}+\textbf{p}_{2} = \textbf{p}_{R}+\textbf{p}^{\prime}$.
For example, the $\Delta$(1232) optical potential is calculated via the nucleon optical potential by \cite{Li23}
 \begin{eqnarray}
  && U_{\Delta ^{- } }  =U_{n}, \quad   U_{\Delta ^{++ } }  =U_{p},  \quad
  U_{\Delta ^{+ }} = \frac{1}{3} U_{n} + \frac{2}{3}U_{p},      \nonumber \\
  &&  U_{\Delta ^{0 } }  =\frac{1}{3} U_{p}+  \frac{2}{3}U_{n},
\end{eqnarray}
 where the $U_{n}$ and $U_{p}$ are the single-particle potentials for neutron and proton, respectively. The N$^{\ast}$-nucleon potential is taken as the same with the nucleon-nucleon potential. The density, isospin and momentum dependent single-nucleon potential is obtained as follows
\begin{eqnarray}
U_{\tau}(\rho,\delta,\textbf{p}) && = \alpha\left(\frac{\rho}{\rho_{0}}\right) + \beta \left(\frac{\rho}{\rho_{0}}\right)^{\gamma} + E_{sym}^{loc}(\rho)\delta^{2}        \nonumber \\
&&  +  \frac{\partial E_{sym}^{loc}(\rho)}{\partial\rho}\rho\delta^{2} + E_{sym}^{loc}(\rho)\rho\frac{\partial\delta^{2}}{\partial\rho_{\tau}}   \nonumber \\
&&  + \frac{1}{\rho_{0}}C_{\tau,\tau} \int d\textbf{p}' f_{\tau}(\textbf{r},\textbf{p})[\ln(\epsilon(\textbf{p}-\textbf{p}')^{2}+1)]^{2}         \nonumber \\
&&  + \frac{1}{\rho_{0}}C_{\tau,\tau'} \int d\textbf{p}' f_{\tau'}(\textbf{r},\textbf{p})      \nonumber \\
&&  \times [\ln(\epsilon(\textbf{p}-\textbf{p}')^{2}+1)]^{2}.
\end{eqnarray}
Here $\tau\neq\tau'$, $\partial\delta^{2}/\partial\rho_{n}=4\delta\rho_{p}/\rho^{2}$ and $\partial\delta^{2}/\partial\rho_{p}=-4\delta\rho_{n}/\rho^{2}$. The nucleon effective (Landau) mass in nuclear matter of isospin asymmetry $\delta=(\rho_{n}-\rho_{p})/(\rho_{n}+\rho_{p})$ with $\rho_{n}$ and $\rho_{p}$ being the neutron and proton density, respectively, is calculated through the potential as $m_{\tau}^{\ast}=m_{\tau}/ \left(1+\frac{m_{\tau}}{|\textbf{p}|}|\frac{dU_{\tau}}{d\textbf{p}}|\right)$ with the free mass $m_{\tau}$ at Fermi momentum $\textbf{p}=\textbf{p}_{F}$. The parameters $\alpha$, $\beta$, $\gamma$ and $\rho_{0}$ are set to be the values of -215.7 MeV, 142.4 MeV, 1.322 and 0.16 fm$^{-3}$, respectively. The $C_{\tau,\tau}=C_{mom}(1+x)$, $C_{\tau,\tau'}=C_{mom}(1-x)$ ($\tau\neq\tau'$) and the isospin symbols $\tau$($\tau'$) represent proton or neutron. The values of 1.76 MeV, 5$\times 10^{-4}$ c$^{2}$/MeV$^{2}$ are taken for the $C_{mom}$ and $\epsilon$, respectively, which result in the effective mass $m^{\ast}/m$=0.75 in nuclear medium at saturation density for symmetric nuclear matter. The parameter $x$ as the strength of the isospin splitting with the value of -0.65 is taken in this work, which has the mass splitting of $m^{\ast}_{n}>m^{\ast}_{p}$ in nuclear medium. A imcompressibility modulus of K=230 MeV for isospin symmetric nuclear matter is obtained at the saturation density. Recently, the influence of the pion potential on pion dynamics in heavy-ion collisions has been extensively investigated with different transport models \cite{Ho14,Fe15,Gu15,So15,Co16,Zh17,Fe17}.

\section{III. Results and discussion}

The isospin diffusion in heavy-ion collisions is associated with the gradient of isospin density ($\rho_{n}-\rho_{p}$) and of symmetry potential $U_{sym}=(U_{n}(\rho,\delta,\textbf{p})-U_{p}(\rho,\delta,\textbf{p}))/2\delta$, in which the symmetry energy plays a significant role on the rearrangement of nucleons in the nuclear evolution. The neutron and proton density distributions vary with the evolution of reaction system and neutron/proton ratio is diverse in the different density range. To extract the high-density symmetry energy, the observables emitted from the high-density domain in nuclear collisions are expected. The neutron/proton ratio is a direct probe for extracting the high-density symmetry energy. But detecting the neutrons with high statistics in experiments is still a difficulty task. We calculated the kinetic energy spectra of neutron/proton (N/Z) ratio of free nucleons produced in collisions of $^{132}$Sn+$^{124}$Sn and $^{108}$Sn + $^{112}$Sn at 270\emph{A} MeV with the different stiffness of symmetry energy as shown in Fig. 2. The free nucleons are judged via the minimum spanning tree (MST) procedure with the relative distance $r_{0}=3$ fm and relative momentum $p_{0}=200$ MeV/c in phase space \cite{Fe11}. It is obvious that the hard symmetry energy with L=139 MeV leads to the larger N/Z ratio, in particular at the energies above 100 MeV for the neutron-rich system. The results are consistent with the conclusion in Fig. 1 (b), which is caused from the more repulsive interaction for neutrons in the dense neutron-rich matter. More free neutrons are created with the hard symmetry energy. Hunting the measurable quantities for constraining the high-density symmetry energy is very necessary, e.g., $\pi^{-}/\pi^{+}$, K$^{0}$/K$^{+}$, $\Sigma^{-}/\Sigma^{+}$ etc.

\begin{figure*}
\includegraphics[height=8cm,width=15cm]{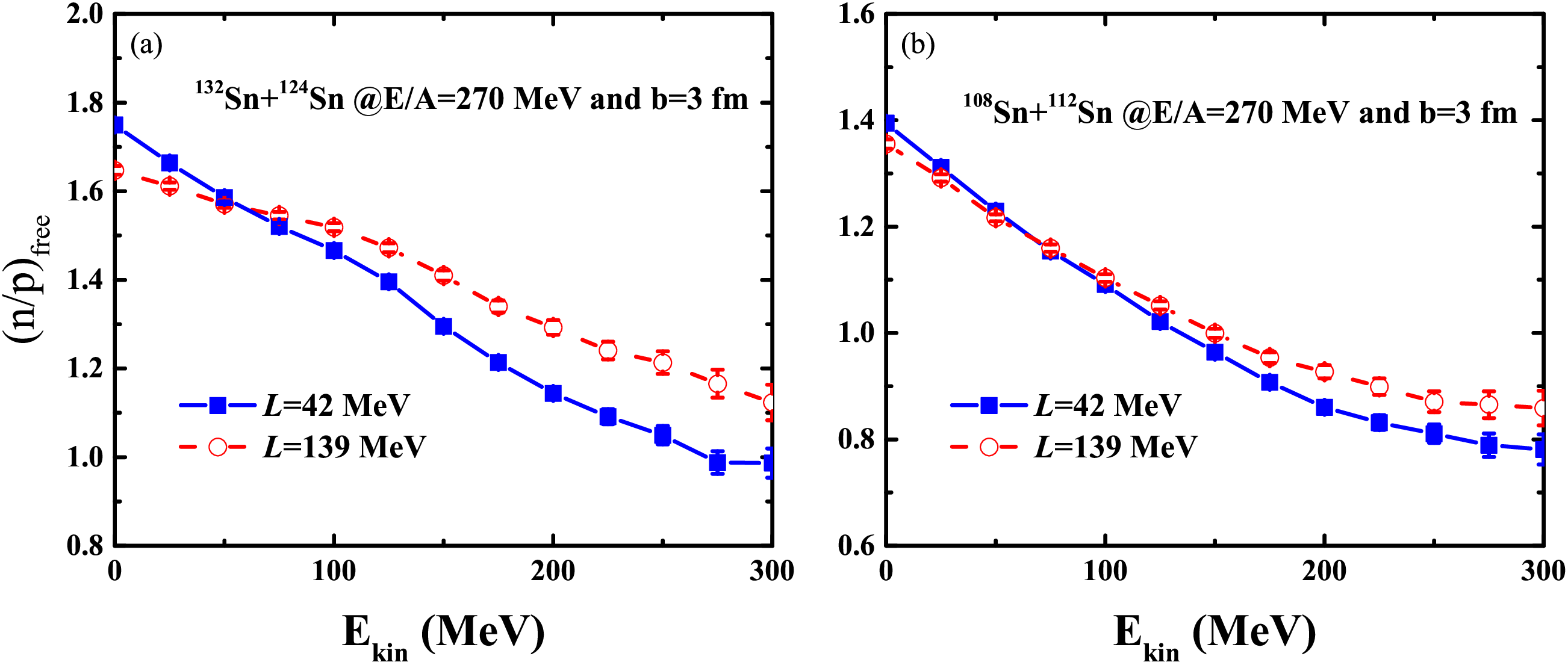}
\caption{ The kinetic energy spectra of free neutron/proton ratio produced in collisions of $^{132}$Sn+$^{124}$Sn (left panel) and $^{108}$Sn+$^{112}$Sn (right panel), respectively. }
\label{fig.2}
\end{figure*}

\begin{figure}
    \centering
 \includegraphics[height=6.5cm,width=7cm]{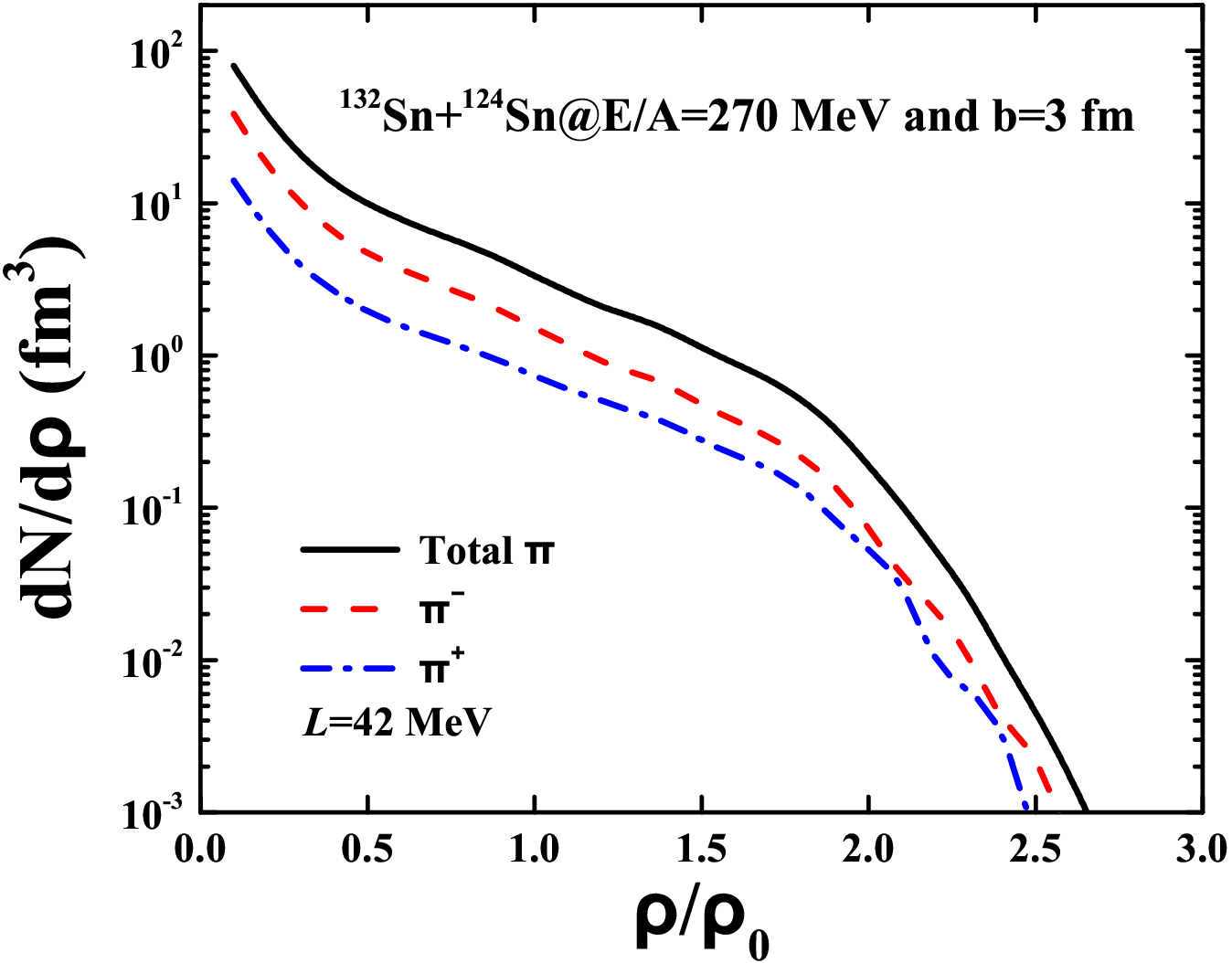}
 \caption{ The density profile of pion production in collisions of $^{132}$Sn+$^{124}$Sn at 270\emph{A} MeV.}
 \label{fig.3}
\end{figure}

\begin{figure*}
   \centering
    \includegraphics[height=6.5cm,width=7cm]{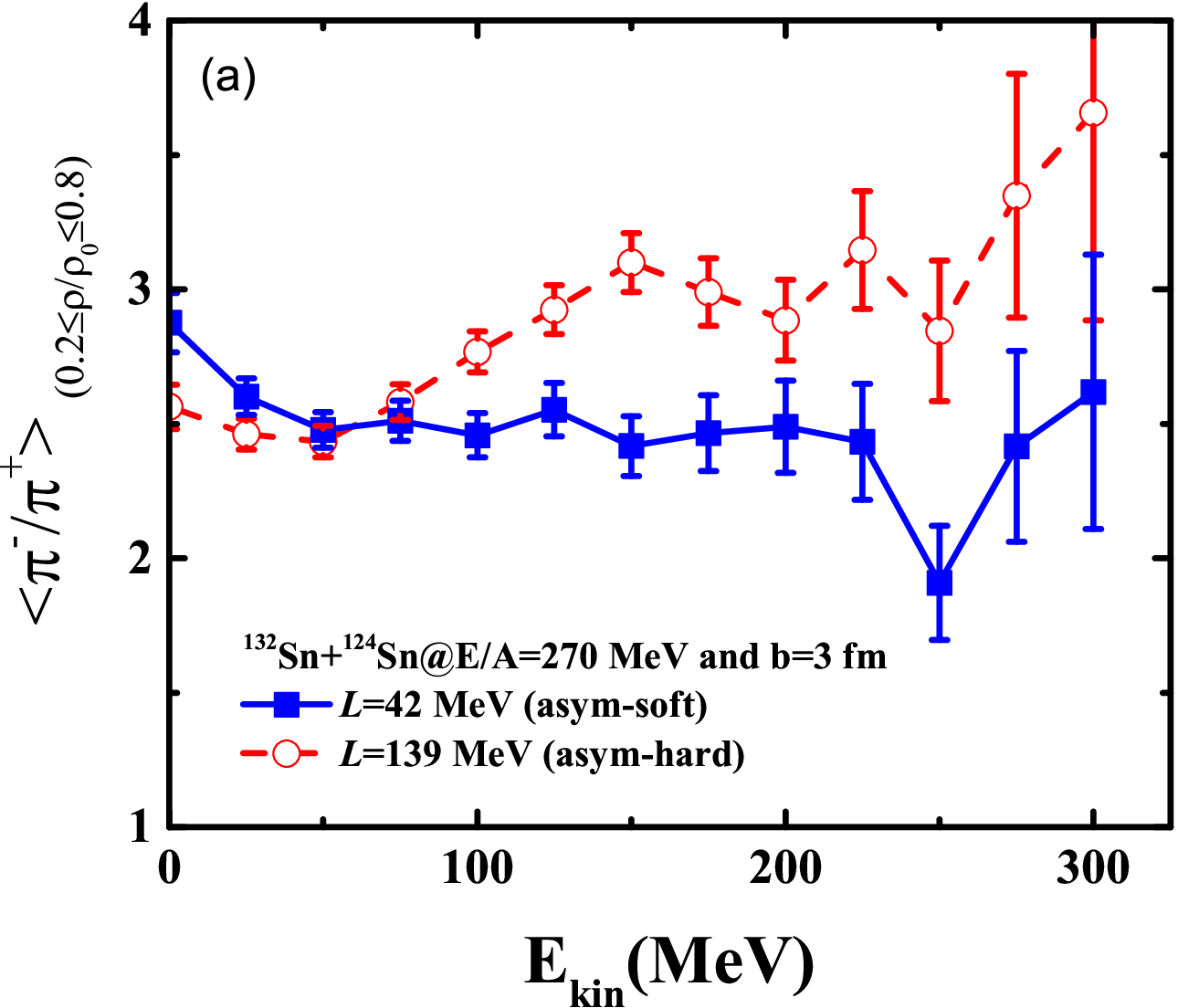}    \includegraphics[height=6.5cm,width=7cm]{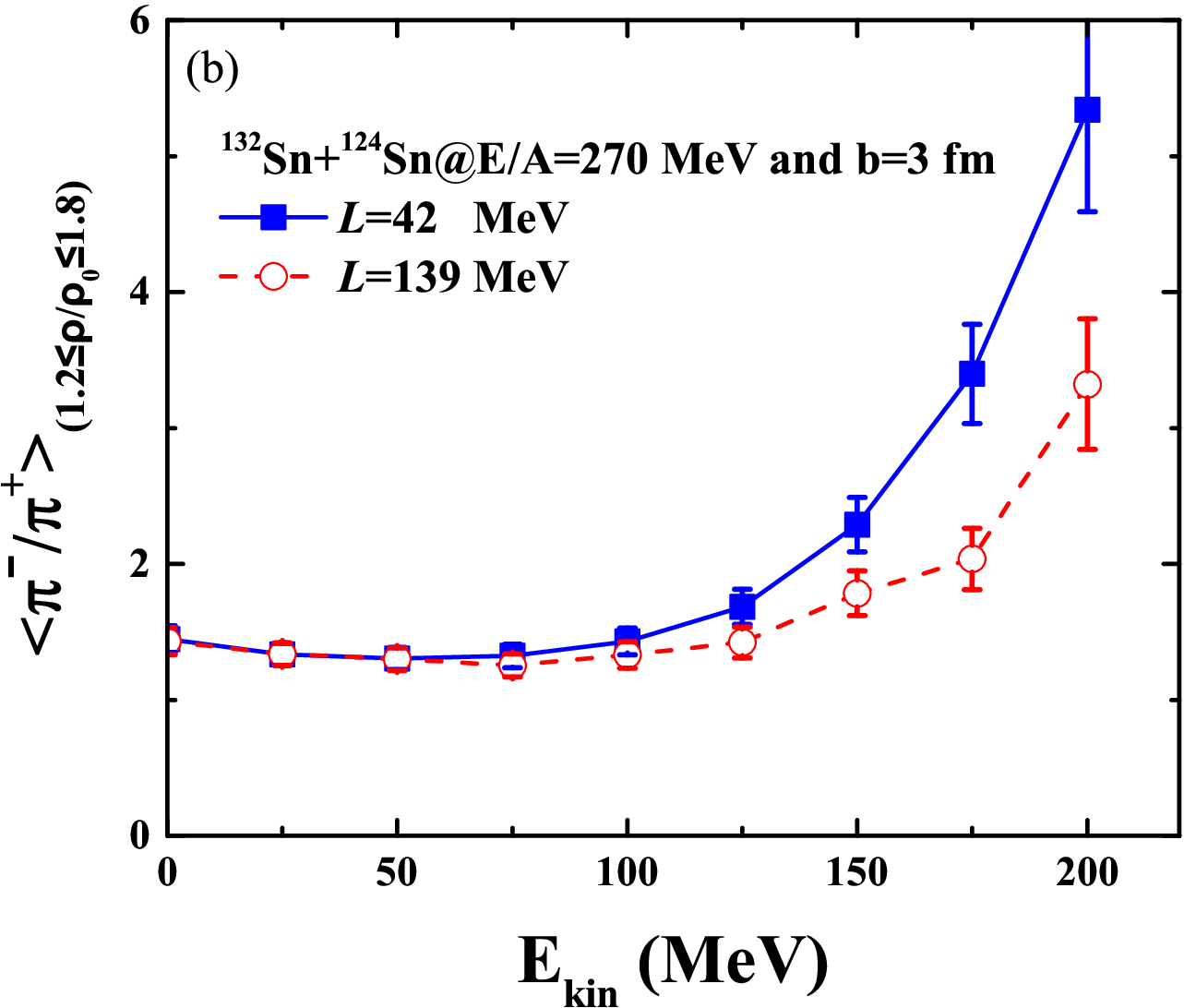}
    \caption{ The kinetic energy spectra of $\pi^{-}/\pi^{+}$ ratio in collisions of $^{132}$Sn+$^{124}$Sn at 270\emph{A} MeV within the baryon density region for the pion production of (a) 0.2$\leq\rho/\rho_{0}\leq$0.8 and (b) 1.2$\leq\rho/\rho_{0}\leq$1.8, respectively. }
 \label{fig.4}
\end{figure*}

The symmetry energy as a major ingredient in the nuclear equation of state plays a significant role on the mass-radius relation and maximal mass of neutron stars, phase transition from quark-glue plasma to hadron phase, nuclear structure of rare isotopes etc. The high-density behavior of symmetry energy from the pion production in heavy-ion collisions has been extensively investigated via transport models, such as the isospin dependent Boltzmann-Uehling-Uhlenbeck (IBUU) and relativistic Vlasov-Uehling-Uhlenbeck (RVUU) \cite{Fe06,Zh17,Yo21,Ga13}. Recently, the transverse momentum spectra of pions has been measured by the S$\pi$RIT collaboration \cite{Es21,Jh21}. In comparison with the previous works, the energy conservation is considered in the decay of resonances. The density profile of primordial pion production in collisions of $^{132}$Sn + $^{124}$Sn at 270\emph{A} MeV is shown as in Fig. \ref{fig.3}. The black line denotes the total pion production ($\pi^{-}+\pi^{0}+\pi^{+}$). And the red dashed and the blue dotted-dashed lines represent $\pi^{-}$ and $\pi^{+}$, respectively. It is obvious that a number of pions are produced in the low-density region, but pions at the suprasaturation densities are still appreciable. The low-density pions is mainly caused from the rescattering processes, i.e., $\pi N\leftrightarrow\Delta(1232)$ and $\Delta(1232)N\leftrightarrow NN$ \cite{Fe17}. Therefore, to extract the high-density information of symmetry energy, one needs to distinguish the final pions from the different density range. It is known that the $\pi^{-}/\pi^{+}$ ratio of primordial pions without the rescatterings is approximately satisfied to the quadratic relation $\pi^{-} /\pi^{+}=\frac{5n^{2}+np}{5p^{2}+np}\approx \left ( n/p \right )^{2}$ \cite{Br75}. To clarify the pion production in the different density range, shown in Fig. \ref{fig.4} is the kinetic energy spectra of $\pi^{-}/\pi^{+}$ ratio in the density region of 0.2$\leq\rho/\rho_{0}\leq$0.8 and 1.2$\leq\rho/\rho_{0}\leq$1.8, respectively. It is pronounced that the symmetry energy effect is opposite in the different density domain, namely, the larger $\pi^{-}/\pi^{+}$ ratio with the hard symmetry energy in the low-density region, but the lower $\pi^{-}/\pi^{+}$ value at high densities. Moreover, the average $\pi^{-}/\pi^{+}$ ratio in the low-density region of 0.2$\leq\rho/\rho_{0}\leq$0.8 is close to the square of N/Z ratio (2.43) of reaction system. But the $\pi^{-}/\pi^{+}$ ratio in the high-density region is approximately the average N/Z value (1.56) at the kinetic energies below 100 MeV. Therefore, the $\pi^{-}/\pi^{+}$ ratio from the total pion multiplicities is indistinguishable for extracting the high-density symmetry energy. The density profile of pions is influenced by the rescattering cross sections and pion evolution in nuclear medium, which might be different in transport models although the total multiplicity is similar. The phase-space distribution of pion is associated with its emission local density. The kinetic energy or transverse momentum spectra of $\pi^{-}/\pi^{+}$ ratio might be available probes for constraining the high-density symmetry energy.

\begin{figure*}
\includegraphics[height=12cm,width=16cm]{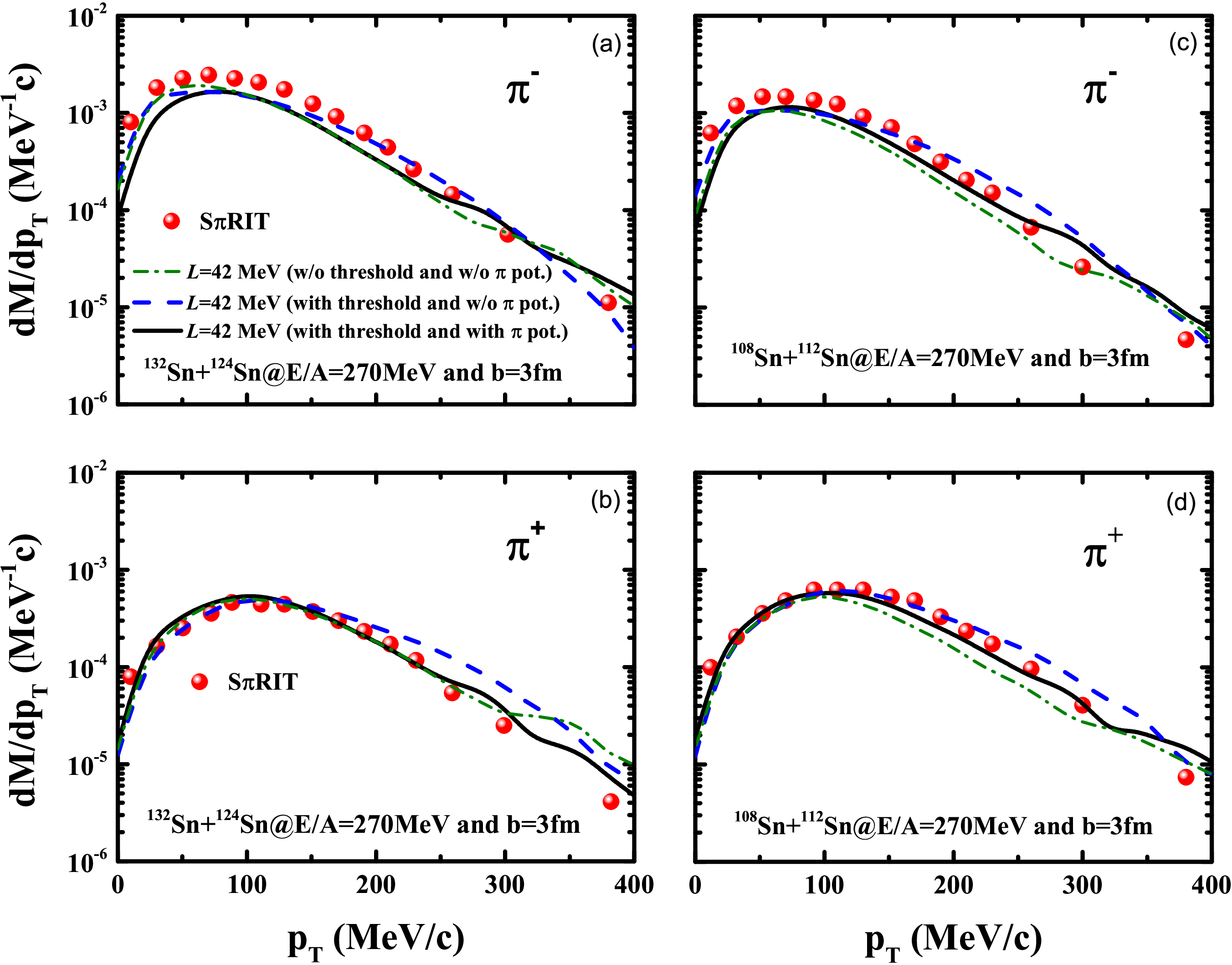}
\caption{Comparison of the transverse momentum spectra with the inclusion of threshold energy correction and pion potential in collisions of $^{132}$Sn+$^{124}$Sn (left panel) and $^{108}$Sn+$^{112}$Sn (right panel), respectively. The experimental data are take from the S$\pi$RIT collaboration \cite{Es21}. }
\label{fig.5}
\end{figure*}

\begin{figure}
\includegraphics[height=7cm, width=9cm] {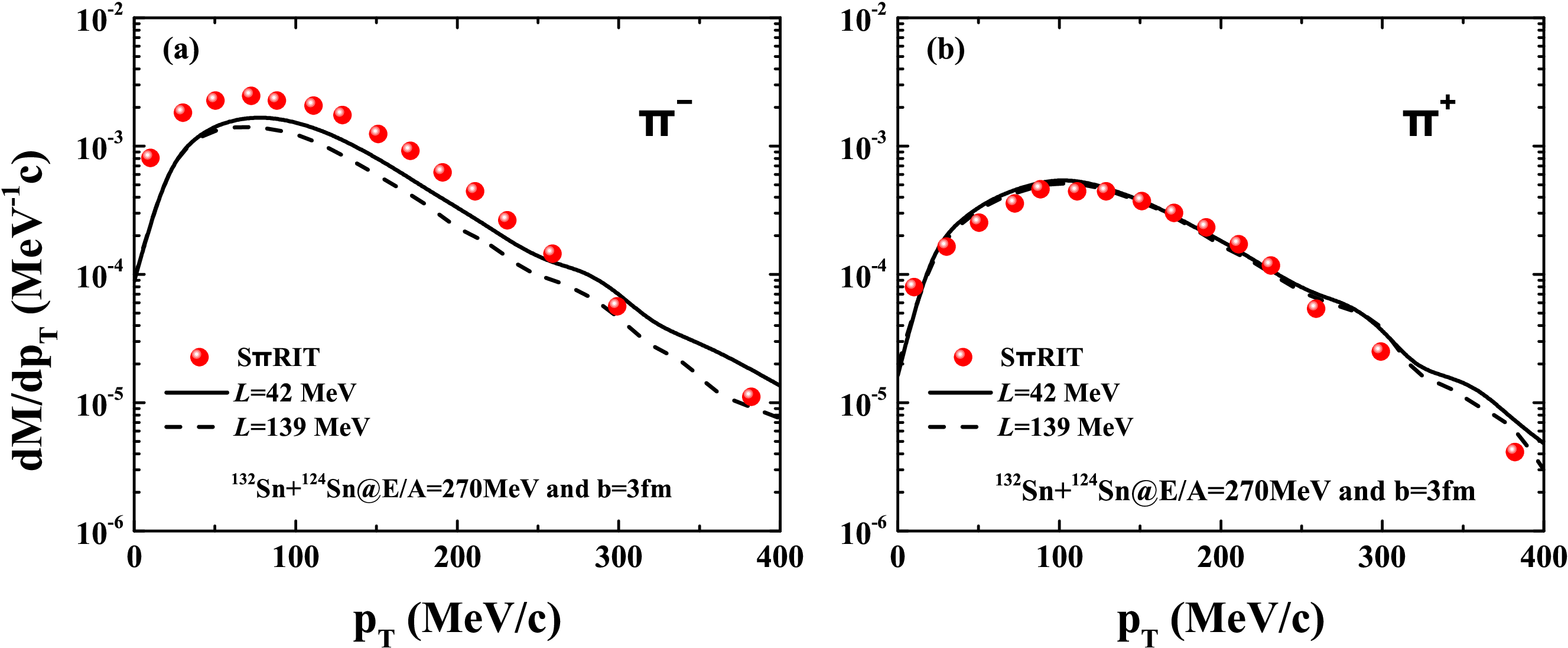}
\caption{ The transverse momentum distribution of pion production with different symmetry energy in the reaction of $^{132}$Sn+$^{124}$Sn at the beam energy of 270 MeV/nucleon and collision centrality of 3 fm and compared with the available data from S$\pi$RIT collaboration \cite{Es21}. }
\label{fig.6}
\end{figure}

The phase-space distribution of emitted particles in heavy-ion collisions manifest the in-medium properties, i.e., the rapidity, transverse momentum, kinetic energy, invariant mass spectra etc. More information of the pion production can be obtained from the transverse momentum spectra. Shown in Fig. \ref{fig.5} is a comparison of the pion potential and energy conservation in the resonance production and in the decay of resonance on the final pion production in collisions of $^{132}$Sn+$^{124}$Sn and $^{108}$Sn+$^{112}$Sn at the incident energy of 270\emph{A} MeV with the soft symmetry energy. The correction of energy conservation includes the optical potentials of resonances and nucleons, pion potentials and Coulomb interaction for the charged pair creation. It is obvious that the inclusion of the energy conservation with Eqs (7) and (8) enhances the high-momentum pion production for both systems, which is caused from the fact that the energetic pions are created for satisfying the energy conservation in the resonance decay in Eq. (7) for the case with threshold correction and without $\pi$ potential (blue dashed line). The term $V_{\pi N}^{Coul}$ in Eq. (7) has the negligible contribution on the pion transverse momentum spectra by the Coulomb interaction between $\pi^{-}$ and proton or $\pi^{+}$ and proton from the resonance decay. The pion-nucleon potential leads to the reduction of pion yields because of the attractive interaction, in particular for the $\pi^{+}$ production. The experimental data of transverse momentum spectra from the S$\pi$RIT collaboration \cite{Es21} are nicely reproduced with the inclusion of energy conservation relation and pion potential. The symmetry energy effect is also analyzed as shown in Fig. \ref{fig.6}. The effect is pronounced for the $\pi^{-}$ production and the soft symmetry energy enhances the yields. The $\pi^{+}$ spectra weakly depend on the stiffness of symmetry energy because the proton-proton and proton-neutron collisions are almost not influenced by the symmetry energy in neutron-rich system. It is concluded that $\pi^{-}/\pi^{+}$ ratio is also enhanced with the symmetry energy of L=42 MeV and consistent with the high-density results in Fig. 4(b). The high-transverse momentum pions are mainly produced from the high-density zone in nuclear collisions and might be probes of high-density symmetry energy, i.e. the pion produced above 150 MeV/c.

\begin{figure}
\includegraphics[height=7cm, width=9cm]{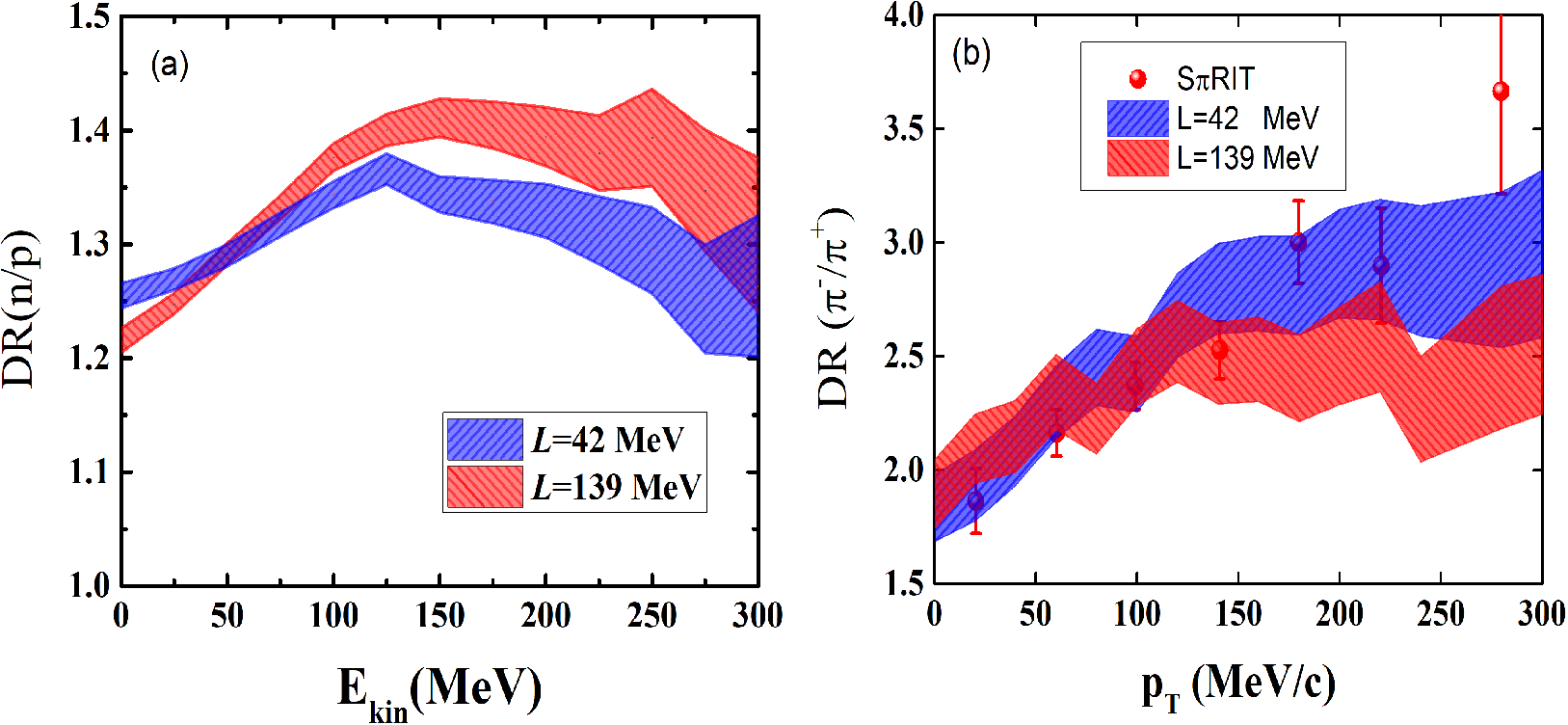}
 \caption{ The double ratios of (a) neutron/proton and (b) $\pi^{-}/\pi^{+}$ in the isotropic reactions of $^{132}$Sn + $^{124}$Sn and $^{108}$Sn + $^{112}$Sn at 270\emph{A} MeV and b= 3 fm. The experimental data for the pion production are taken from the S$\pi$RIT collaboration \cite{Es21}.}
\label{fig.7}
\end{figure}

To eliminate the Coulomb interaction and reduce some uncertainties, we analyzed the double ratios (DRs) in the isotopic reactions, which are defined as the relation DR(n/p)=$(n/p)^{free}_{^{132}Sn + ^{124}Sn}/(n/p)^{free}_{^{108}Sn + ^{112}Sn}$ and DR($\pi^{-}/\pi^{+}$)=$(\pi^{-}/\pi^{+})_{^{132}Sn + ^{124}Sn}/(\pi^{-}/\pi^{+})_{^{108}Sn + ^{112}Sn}$, respectively. The method has been used for constraining the symmetry energy at the subsaturation density and the isospin splitting of nucleon effective mass in nuclear matter via heavy-ion collisions \cite{Gu17}. The symmetry energy effect is pronounced at the high kinetic energies (high transverse momenta) in the isotopic reactions as shown in Fig. \ref{fig.7}, i.e., in the domain of $E_{kin}>$100 MeV or $p_{T}>$150 MeV/c. The double ratio of neutron/proton has more obvious symmetry energy effect in comparison with the case of $\pi^{-}/\pi^{+}$, but the larger DR value for the pion production. The complementary observables manifest that the DRs of high kinetic energy or transverse momentum might be probes for extracting the high-density symmetry energy, i.e., the DR(n/p) and DR($\pi^{-}/\pi^{+}$) spectra. The slope parameter of symmetry energy at saturation density $L(\rho_{0}) = 42\pm 25$ is obtained by using the standard error analysis method within the range $1\sigma$. Due to the mixing of pions produced at high and low densities, the symmetry energy effect on the DR spectra is reduced. However, at high transverse momentum ($p_{T}\ge$ 150 MeV/c) the DRs exhibit significant sensitivity to the high-density symmetry energy, and the difference is about 20$\%$. It is noted that the constraint of 42 MeV $<L<$ 117 MeV is obtained by the dcQMD calculations. Different with the dcQMD model, we fixed the isospin splitting of nucleon effective mass, i.e., $(m_{n}^{\ast}-m_{p}^{\ast})/m_{n}=0.04$ with the isospin asymmetry of $\delta=0.2$ at the saturation density. The calculations from the dcQMD model manifest that the hard symmetry energy with $L(\rho_{0}) = 151$ MeV leads to the larger single ratio $\pi^{-}/\pi^{+}$ and DRs at the high transverse momenta, which is possibly caused from the larger pion-nucleon scattering cross section and the dominating contribution of the low-density pions on the transverse momentum spectra. In comparison with our previous results for constraining the high-density symmetry energy from the pion production in heavy-ion collisions \cite{Fe10a}, the isospin, density and momentum dependent nucleon-nucleon and pion-nucleon potentials, the energy conservation of resonance production and reabsorption with the channel of NN$\leftrightarrow$NR, and the energy conservation of pion production and reabsorption with the channel of R$\leftrightarrow \pi$N have been implemented into the LQMD model. The modifications are essential for the transverse momentum spectra of pions near threshold energies. It is obvious that the symmetry energy effects of DR(n/p) and DR($\pi^{-}/\pi^{+}$) spectra are opposite in the high kinetic energy or transverse momentum region. The larger values of DR(n/p) and lower values of DR($\pi^{-}/\pi^{+}$) with the hard symmetry energy are complementary and manifest that the pions in the high transverse momentum ($p_{T}\ge$ 150 MeV/c) are mainly produced in the high-density domain. The results are also self-consistent with the kinetic energy spectra of the single ratios of $(n/p)_{free}$ and $\pi^{-}/\pi^{+}$ in the high-density domain.

The mass-radius relation of neutron stars (NSs) is influenced by the density dependent symmetry energy and properties of isospin symmetric nuclear matter. It has been investigated that the maximal mass of NSs is related to the imcompressibility modulus of nuclear matter and the stiffness of hyperon-nucleon interaction potential \cite{We24plb}. Shown in Fig. 8 is a comparison of the density dependence of pressure and the mass-radius relation of the neutron matter with the different slope parameter of symmetry energy. The constraint of symmetry energy from the pion production in heavy-ion collisions by the S$\pi$RIT data leads to the maximal mass of 2.0 $M_{\odot}$ and radius of 11-13 km as shown the blue lines in Fig. 8 (b). The strange NS with the mass of 2.76 $M_{\odot}$ and the radius of 15 km is obtained with the hard symmetry energy of L=139 MeV. The results are consistent with the NICER observations for PSR J0030+0451 and J0740+6620. It is obvious that the symmetry energies with the slope parameters of L=35, 42, 53 and 67 MeV result in the same maximal mass of NSs, but for the different radius. The properties of symmetric nuclear matter and hyperon ingredient in the NSs influence the maximal mass.

\begin{figure*}
\includegraphics[height=8cm, width=16cm]{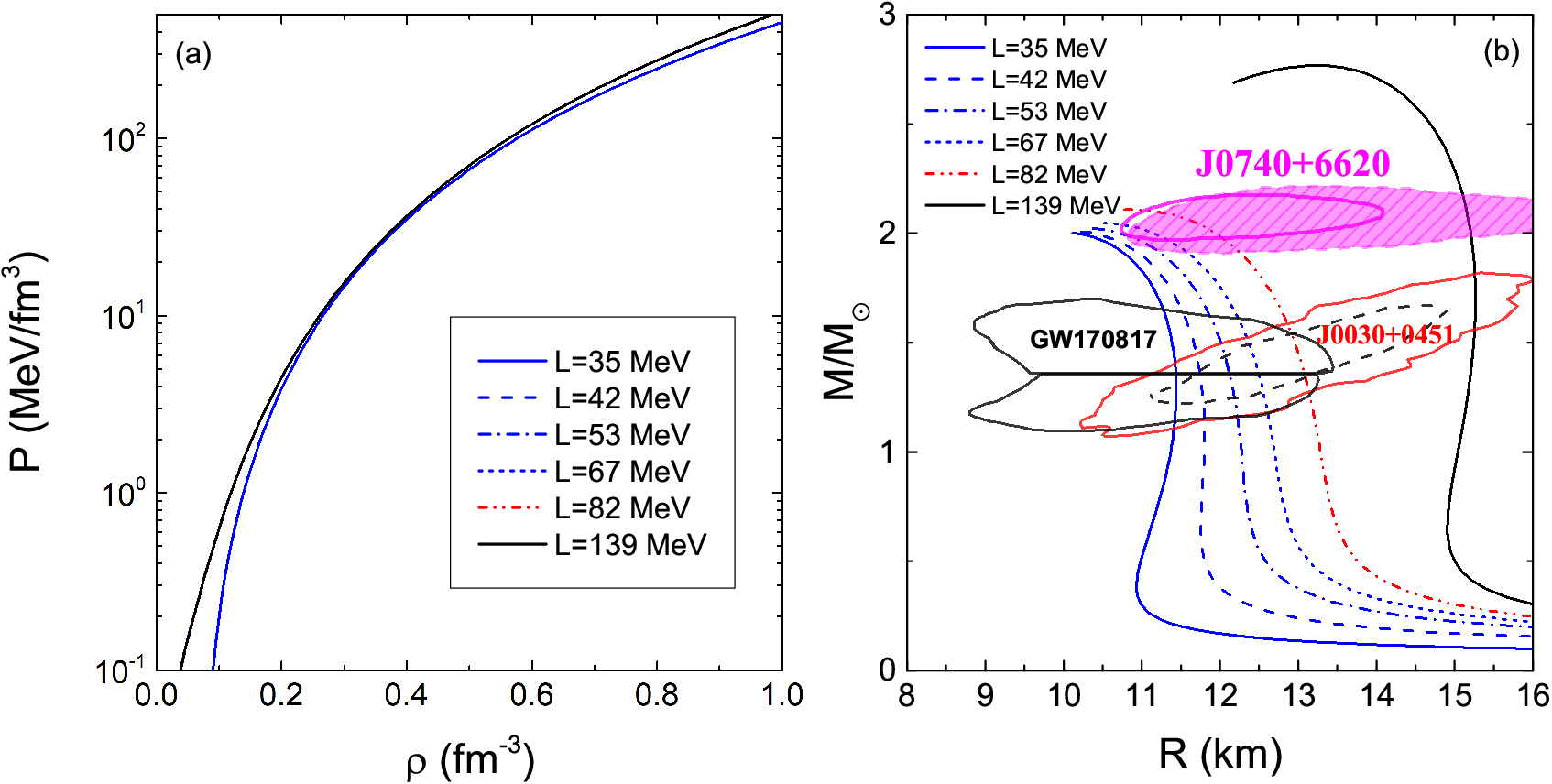}
 \caption{ (a) The pressure as a function of baryon density and (b) mass-radius relation of neutron star matter. The constraints from the GW170817 event \cite{Ab17,Ab18} and from the NICER observations for J0030+0451 \cite{Ri19,Mi19} and J0740+6620 \cite{Cr19,Ri21} are shown for comparison. }
\label{fig.8}
\end{figure*}

\section{IV. Conclusions}
 In summary, the isospin diffusion in the isotopic reactions of $^{132}$Sn + $^{124}$Sn and $^{108}$Sn + $^{112}$Sn at the incident energy of 270 MeV/nucleon is thoroughly investigated within the LQMD transport model. The symmetry energy manifests the opposite contribution on the N/Z and $\pi^{-}/\pi^{+}$ ratios in the low-density region and in the high-density domain. The hard symmetry energy enhances the N/Z ratio of free nucleons, but leads to the reduction of N/Z and $\pi^{-}/\pi^{+}$ ratios in the high-density region. The production of pions in heavy-ion collisions spreads the whole density range and most of pions are created in the dilute nuclear matter because of the rescattering processes of pions and resonances with nucleons. The stiffness of symmetry energy is sensitive to the $\pi^{-}$ transverse spectra, but weakly impact the $\pi^{+}$ production. The pion optimal potential is obvious in the high-momentum regime and influences both the $\pi^{-}$ and $\pi^{+}$ spectra. Systematics analysis with the symmetry energy and pion-nucleon potential at the density around $1.5\rho_{0}$ manifests the soft symmetry energy with the slope parameter $L(\rho_{0}) = 42\pm 25$ MeV or stiffness parameter $\gamma_{s}=$0.3 with the S$\pi$RIT data. More experiments are still expected for complementary constraining the high-density symmetry energy, i.e., the collective flows and differential flows of pions, the energy spectra of triton/$^{3}$He ratio etc. The neutron stars with the maximal mass of 2 $M_{\odot}$ and radius of 11-13 km are obtained with the symmetry energy extracted from the pion data, which are consistent with the NICER observations for PSR J0030+0451 and J0740+6620.

\textbf{Acknowledgements}
This work was supported by the National Natural Science Foundation of China (Projects No. 12175072 and No. 12311540139) and the Talent Program of South China University of Technology (Projects No. 20210115).

\end{document}